\documentclass[onecolumn,useAMS,usenatbib]{mn2e}

\usepackage{graphicx}
\usepackage{subfigure}
\usepackage{xcolor}
\usepackage{mathtext,bbm,amsmath,amsfonts,amssymb,indentfirst,syntonly,graphicx}
\usepackage{mathtools}
\usepackage{slashbox}
\usepackage[english]{babel}
\usepackage{calc}
\usepackage{tikz}

\def\bc{\begin{center}}
\def\ec{\end{center}}
\def\be{\begin{eqnarray}}
\def\ee{\end{eqnarray}}

\title[Comparison between the HC method and DF method]{Comparison between hemisphere comparison method and dipole-fitting method in tracing the anisotropic expansion of the Universe use the Union2 dataset}
\author[Z. Chang and H.-N. Lin]
        {Zhe Chang$^{1,2}$ and Hai-Nan Lin$^{1}$\thanks{E-mail: linhn@ihep.ac.cn.}\\
$^{1}$Institute of High Energy Physics, Chinese Academy of Sciences, 100049 Beijing, China\\
$^{2}$Theoretical Physics Center for Science Facilities, Chinese Academy of Sciences, 100049 Beijing, China}
\begin{document}

\date{Accepted xxxx; Received xxxx; in original form xxxx}

\pagerange{\pageref{firstpage}--\pageref{lastpage}} \pubyear{2014}

\maketitle

\label{firstpage}

\begin{abstract}
Type-Ia supernovae (SNe Ia) are often used as the standard candles to probe the anisotropic expansion of the Universe. In this paper, we make a comprehensive comparison between the hemisphere comparison (HC) method and dipole-fitting (DF) method in searching for the cosmological preferred direction using the Union2 dataset, a compilation of 557 well-calibrated SNe Ia. We find that the directions of the faintest SNe Ia derived from these two methods are approximately opposite. Monte Carlo simulations show that the results of the HC method strongly depend on the distribution of the data points in the sky. The coincidence that the HC method and DF method give two completely opposite directions may be due to the extremely nonuniform distribution of the Union2 dataset.
\end{abstract}


\section{Introduction}\label{sec:introduction}

The cosmological principle, which states that the Universe is homogeneous and isotropic on large scale, is one of the foundations of the Modern cosmology. However, the cosmological principle is strongly doubted recently. Observations on the large-scale structure of the Universe, such as the large-scale bulk flow \citep{Kashlinsky:2008,Watkins:2009,Lavaux:2010}, the alignments of low multipoles in the CMB angular power spectrum \citep{Lineweaver:1996,Tegmark:2003,Bielewicz:2004,Copi:2010,Frommert:2010}, the large-scale alignments of the quasar polarization vectors \citep{Hutsemekers:2005,Hutsemekers:2011}, the spatial variation of the fine-structure constant \citep{King:2012,Mariano:2012}, so on, show that the Universe may be anisotropic. Especially, the recent released WMAP data \citep{Bennett:2011,Bennett:2013} and Planck data \citep{Ade:2013a,Ade:2013b} further arouse great interests in probing the possible anisotropy of the Universe. \citet{Bonvin:2006} analyzed a sample of nearby supernovae and find a dipole of luminosity distance consistent with the CMB at a significance of more than $2\sigma$. More interestingly, it is found that the CMB maximum temperature asymmetry axis is aligned with dark energy dipole and bulk flow directions \citep{Antoniou:2013}. All of these hint that there may exist a preferred direction in the Universe.

Type-Ia supernovae (SNe Ia) are the ideal distance indicators to trace the expansion history of the Universe due to their consistent absolute magnitudes. In fact, they have been widely used as the standard candles to search for the preferred direction of the Universe. Generally speaking, there are two different ways to investigate the possible anisotropy from the data. One is directly fitting the data to a specific anisotropic cosmological model \citep{Campanelli:2011,Li:2013,Chang:2013,Chang:2014a,Chang:2014b,Schcker:2014}. Many anisotropic cosmological models have be proposed to match the observations. The Bianchi I type cosmological model \citep{Campanelli:2011,Schcker:2014} and the Rinders-Finsler cosmological model \citep{Chang:2013,Chang:2014a,Chang:2014b} are two models which are consistent with the SNe Ia data. A scalar perturbation of the $\Lambda$CDM model may also break the spherical symmetry of the Universe such that a preferred axis arises \citep{Li:2013}. \citet{Mariano:2012} proposed the extended topological quintessence model which leads to a spherical inhomogeneous distribution for dark energy density. The large-scale matter density inhomogeneities would naturally introduce accelerating expansion anisotropy of the Universe.

An alternative method is to analyze the SNe Ia data in a model-independent way \citep{Antoniou:2010,Mariano:2012,CaiTuo:2012,Cai:2013,kalus:2013,Zhao:2013,Yang:2014}. The advantage of this method is that it does not rely on the particular cosmological model. The hemisphere comparison (HC) method and dipole-fitting (DF) method are two most used methods in literatures. The HC method divides the data points into two subsets according to their position in the sky and fit the subsets to an isotropic cosmological model (e.g., ${\rm\Lambda}$CDM model) independently, while the DF method directly fits the data points to a dipole (or dipole plus monopole) model. \citet{Antoniou:2010} found that the SNe Ia are the faintest in the direction pointing to $(l,b)=(309^{\circ},18^{\circ})$ in the Union2 dataset using the HC method. The work of \citet{CaiTuo:2012} further confirms this result. In a later paper, \citet{Mariano:2012} found a dipole direction pointing towards $(l,b)=(309.4^{\circ},-15.1^{\circ})$ in the same dataset, but using the DF method. At the first glance, these two preferred directions are consistent within uncertainties. However, further investigation shows that the dipole direction corresponds to the brightest SNe Ia, while the faintest SNe Ia are in the opposite direction (see the next section for details). This is to say, the result of the HC method completely conflicts with that of the DF method. This triggers us to further study the discrepancies between these two methods.

This paper aims to addressing the above issues. The rest of the paper is organized as follows: In section \ref{sec:comparison}, we make a comprehensive comparison between the HC method and DF method using the Union2 compilation. It is found that there are obvious discrepancies between these two methods. In section \ref{sec:discussion}, we discuss the reasons that may cause the discrepancies. Finally, section \ref{sec:summary} is devoted to a short summary.

\section{Comparison between the HC method and DF method}\label{sec:comparison}

In this section, we make a comprehensive comparison between the HC method and the DF method using the Union2 dataset. This section is divided into four subsections. In section \ref{sec:dataset}, we introduce the Union2 dataset that is used in the analysis and the basic formulae of the ${\rm \Lambda}$CDM model. Section \ref{sec:hemisphere} shortly reviews the HC method, while section \ref{sec:dipole} describes the DF method. We find that the directions of the faintest SNe Ia obtained using these two methods are approximately opposite. In section \ref{sec:simulation}, we apply Monte Carlo simulations to further investigate the discrepancies between these two methods.

\subsection{The Union2 dataset and ${\rm \Lambda}$CDM model}\label{sec:dataset}

\begin{figure}
\centering
  \includegraphics[width=16 cm]{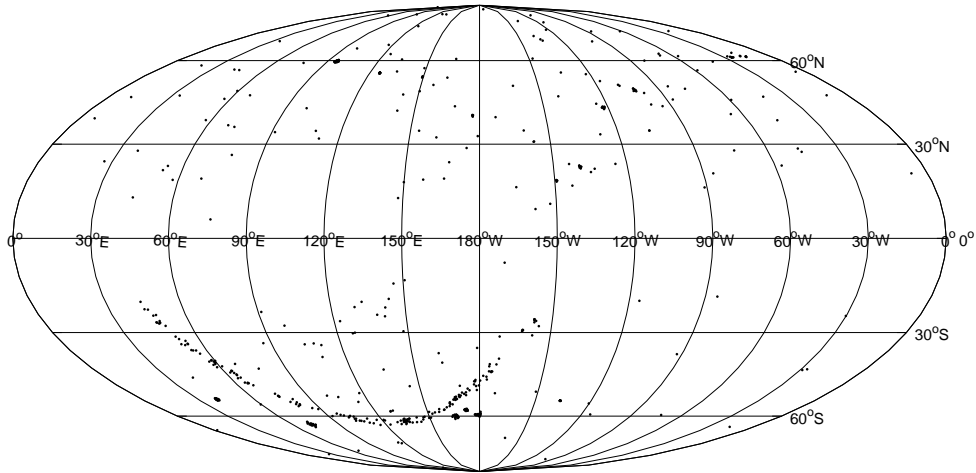}
  \vspace{-1.5cm}
  \caption{\small{The distribution of the Union2 data points in the galactic coordinates system.}}\label{fig:distribution}
\end{figure}

The Union2 dataset \citep{Amanullah:2010} consists of 557 SNe Ia with well-observed redshift in the range of $z\in[0.015,1.4]$. The distance moduli and their uncertainties are extracted from the SALT2 light-curve fitter. The directions of SNe Ia are well localized in the sky of the equatorial coordinates system \citep{Blomqvist:2010}, which can be easily converted to the galactic coordinates system \citep{Smith:1989}. The distribution of the Union2 dataset in the sky of the galactic coordinates system is showed in Fig.\ref{fig:distribution}.
As can be seen, the SNe Ia are not uniformly distributed. A lot of data points cluster near a line corresponding to the equator of the equatorial coordinates system.

We work in the background of the standard cosmological model, i.e., the ${\rm \Lambda}$CDM model, in which the luminosity distance is written as
\begin{equation}\label{eq:luminosity-distance}
  D_L(z)=(1+z)\frac{c}{H_0}\int_0^z \frac{dz}{\sqrt{\Omega_M(1+z)^3+1-\Omega_M}},
\end{equation}
where $c$ is the speed of light, $H_0$ is the Hubble constant, and $\Omega_M$ is the normalized matter density. The luminosity distance can be conveniently converted to the dimensionless distance modulus as
\begin{equation}\label{eq:distance-modulus}
  \mu(z)=5\log_{10}\frac{D_L(z)}{\rm Mpc}+25.
\end{equation}

The best-fit parameters ($\Omega_M,H_0$) can be obtained using the usual least-$\chi^2$ method. Define $\chi^2$ as
\begin{equation}
  \chi^2=\sum_{i=1}^N\left(\frac{\mu_{\rm th}^{(i)}-\mu_{\rm obs}^{(i)}}{\sigma_{\mu}^{(i)}}\right)^2,
\end{equation}
where $N=557$ is the number of SNe Ia in the Union2 dataset, $\mu_{\rm th}$ is theoretical distance modulus calculated from Eqs.(\ref{eq:luminosity-distance}) and (\ref{eq:distance-modulus}), $\mu_{\rm obs}$ is the observed distance modulus, and $\sigma_{\mu}$ is the uncertainty of $\mu_{\rm obs}$. The sum runs over all the data points. The best-fit values are the ones which can minimize $\chi^2$, i.e.,
\begin{equation}\label{eq:parameter}
  \Omega_M=0.27\pm 0.02,~~~~H_0=70.0\pm 0.4~~{\rm km}~{\rm s}^{-1}~{\rm Mpc}^{-1}.
\end{equation}

\subsection{The HC method}\label{sec:hemisphere}

To investigate the possible anisotropic expansion of the Universe, we apply the HC method to the Union2 dataset. The main procedures of the HC method are as follows:
\begin{enumerate}
  \item{Generate a random direction $(l,b)$ in the northern hemisphere of the galactic coordinates system, where $l\in[0^{\circ}, 360^{\circ})$ and $b\in[0^{\circ},90^{\circ}]$ are random numbers following the uniform distribution.}
  \item{The ``\,equator" corresponding to the random direction divides the sky in to two hemispheres, which are called ``\,up" hemisphere and ``\,down" hemisphere, respectively. Correspondingly, the data points are divided into two subsets according to their position in the sky.}
  \item{Find the best-fit parameters ($\Omega_M, H_0)$ in each hemisphere. Define the magnitude of anisotropy in that direction as
      \begin{equation}
        D_{\Omega}(l,b)\equiv\frac{\Delta\Omega_M}{\bar{\Omega}_M}=2\frac{\Omega_{M,u}-\Omega_{M,d}}{\Omega_{M,u}+\Omega_{M,d}},
      \end{equation}
      where $\Omega_{M,u}$ and $\Omega_{M,d}$ are the best-fit parameters in the ``\,up" hemisphere and ``\,down" hemisphere, respectively.}
  \item{Repeat steps (i) to (iii) $n$ times (for example, $n=400$), and find the maximum value of $|D_{\Omega}|$ and the corresponding direction ($l,b$).}
\end{enumerate}

\begin{figure}
\centering
  \includegraphics[width=16 cm]{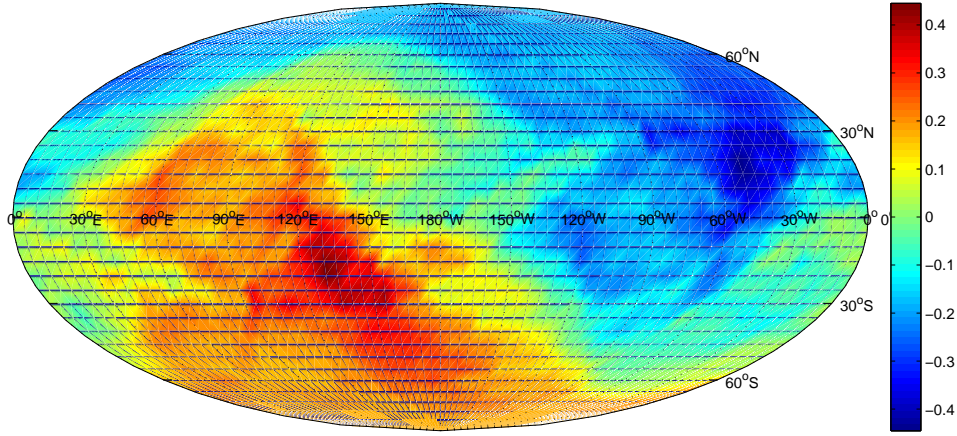}
  \vspace{-1.5cm}
  \caption{\small{The pseudo color map of $D_{\Omega}(l,b)\equiv\Delta\Omega_M/\bar{\Omega}_M$ derived from the HC method. The faintest SNe Ia are in the direction points towards $(l,b)=(309^{\circ}, 18^{\circ})$, while the brightest SNe Ia are in the opposite direction.}}\label{fig:omegam}
\end{figure}

We plot the pseudo color map of $D_{\Omega}(l,b)$ in Fig.\ref{fig:omegam}.
When plotting this figure, instead of choosing random axes, as is described in step (i), we divide the sky into $5^{\circ}\times 5^{\circ}$ grids. The axes are chosen to be towards the center of each grid. This makes the figure to be more smooth, but makes the computation much more time-consuming. The maximum value of $|D_{\Omega}|$ is found to be in the direction \citep{Antoniou:2010}
\begin{equation}
  (l,b)=(309^{\circ}\,_{-3^{\circ}}^{+23^{\circ}},18^{\circ}\,_{-10^{\circ}}^{+11^{\circ}}),~~~~\Omega_M=0.19\pm 0.04,
\end{equation}
or equivalently, its opposite direction\footnote{The $1\sigma$ error of the longitude of the opposite direction given in \citet{Antoniou:2010} is incorrect, where the authors give as $l=129^{\circ}\,_{-23^{\circ}}^{+3^{\circ}}$. Since the opposite direction to $(l,b)$ is $(l\pm 180^{\circ},-b)$, the uncertainty of the longitude of a certain direction is the same to that of its opposite direction.}
\begin{equation}
  (l,b)=(129^{\circ}\,_{-3^{\circ}}^{+23^{\circ}},-18^{\circ}\,_{-11^{\circ}}^{+10^{\circ}}),~~~~\Omega_M=0.30\pm 0.02.
\end{equation}
The $1\sigma$ errors are derived using the method described in \citet{Antoniou:2010}. In this two directions, the magnitude of anisotropy reaches its maximum, i.e., $|D_{\Omega}|_{\rm max}=0.43\pm 0.06$.

To see the possible anisotropy of the Hubble parameter, we apply the anisotropy analysis done for the parameter $\Omega_M$ to the parameter $H_0$. Define the magnitude of anisotropy of the Hubble parameter as
\begin{equation}
        D_H(l,b)\equiv\frac{\Delta H_0}{\bar{H}_0}=2\frac{H_{0,u}-H_{0,d}}{H_{0,u}+H_{0,d}}.
\end{equation}
Following the procedures of the HC method, we find that the maximum value of $H_0$ is in the direction
\begin{equation}
  (l,b)=(334^{\circ}\,_{-6^{\circ}}^{+6^{\circ}},30^{\circ}\,_{-8^{\circ}}^{+2^{\circ}}),~~~~ H_0=71.4\pm 0.6~{\rm km~s}^{-1}~{\rm Mpc}^{-1},
\end{equation}
while the minimum value of $H_0$ is in the opposite direction, i.e.,
\begin{equation}
  (l,b)=(154^{\circ}\,_{-6^{\circ}}^{+6^{\circ}},-30^{\circ}\,_{-2^{\circ}}^{+8^{\circ}}),~~~~ H_0=69.3\pm 0.4~{\rm km~s}^{-1}~{\rm Mpc}^{-1}.
\end{equation}
Therefore, the maximum anisotropy of the Hubble parameter is $|D_{H}|_{\rm max}=0.03\pm 0.01$. This is consistent to the result of \citet{kalus:2013}, who obtained $\Delta H/H<0.038$ at 95\% confidence level using four groups of SNe Ia datasets of redshift $z<0.2$, although the redshifts of the Union2 dataset extend to $z=1.4$. This implies that $H_0$ is not sensitive to the directions. In fact, if we fix $H_0$ to $70.0~{\rm km~s}^{-1}~{\rm Mpc}^{-1}$ (the best-fit value of the $\Lambda$CDM model), and implement the HC method to search the anisotropy of $\Omega_M$, we find a similar direction of maximum anisotropy, pointing towards $(l,b)=(320^{\circ}\,_{-27^{\circ}}^{+8^{\circ}},29^{\circ}\,_{-8^{\circ}}^{+5^{\circ}})$. If we assume that the Hubble parameter is isotropic (this is approximately true, as is showed here) and the absolute magnitude of SNe Ia is a constant, smaller $\Omega_M$ leads to larger luminosity distance so fainter supernova (at a fixed redshift). Therefore, the hemisphere of the faintest SNe Ia is in the direction $(l,b)=(309^{\circ}\,_{-3^{\circ}}^{+23^{\circ}},18^{\circ}\,_{-10^{\circ}}^{+11^{\circ}})$, while the hemisphere of the brightest SNe Ia is in the opposite direction.

\subsection{The DF method}\label{sec:dipole}

To check the reasonableness of the HC method, we apply the DF method to the same dataset. The DF method is a method of directly fitting the data to a dipole (or dipole plus monopole) model. If the Universe is really intrinsically anisotropic and there exists a preferred direction, these two methods should give the similar results. The main steps of the DF method are as follows:
\begin{enumerate}
  \item{Calculate the deviation of the distance modulus of each supernova from its best fit ${\rm \Lambda}$CDM value, i.e., define
  \begin{equation}\label{eq:deviation-of-modulus}
    \frac{\Delta\mu(z)}{\bar{\mu}(z)}\equiv\frac{\bar{\mu}(z)-\mu(z)}{\bar{\mu}(z)},
  \end{equation}
  where $\bar{\mu}$ is the distance modulus predicted by the ${\rm\Lambda}$CDM model with the best-fit parameters given in Eq.(\ref{eq:parameter}), and $\mu$ is the observed distance modulus.}
  \item{Given any axis $\mathbf{\hat{n}}=\cos(b)\cos(l)\mathbf{\hat{i}}+\cos(b)\sin(l)\mathbf{\hat{j}}+\sin(b)\mathbf{\hat{k}}$ in the sky, calculate the angle of each supernova with respect to the axis, which is determined by $\cos\theta_i=\mathbf{\hat{n}}\cdot\mathbf{\hat{p_i}}$, where $\mathbf{\hat{p_i}}=\cos(b_i)\cos(l_i)\mathbf{\hat{i}}+\cos(b_i)\sin(l_i)\mathbf{\hat{j}}+\sin(b_i)\mathbf{\hat{k}}$ is the unit vector pointing towards the $i$-th supernova.}
  \item{Rewrite Eq.(\ref{eq:deviation-of-modulus}) into the dipole plus monopole form, i.e.,
  \begin{equation}\label{eq:dipole}
    \left(\frac{\Delta\mu}{\bar{\mu}}\right)_i=A\cos\theta_i+B,
  \end{equation}
  where $A$ and $B$ are the magnitudes of the dipole and monopole, respectively. Using the least-$\chi^2$ method, we can find the best-fit parameters ($A,B,l,b$) which can minimize the $\chi^2$,
  \begin{equation}
    \chi^2=\sum_{i=1}^N\frac{\left[\left(\frac{\Delta\mu}{\bar{\mu}}\right)_i-A\cos\theta_i-B\right]^2}{\sigma_i^2},
  \end{equation}}
  where $\sigma_i=\sigma_{\mu}^{(i)}/\bar{\mu}^{(i)}$.
\end{enumerate}

The best-fit dipole direction is found to be towards \citep{Mariano:2012}
\begin{equation}\label{eq:direccion1}
  (l,b)=(309.4^{\circ}\pm 18.0^{\circ},-15.1^{\circ}\pm 11.5^{\circ}),
\end{equation}
and the magnitudes of the dipole and monopole are given as
\begin{equation}\label{eq:magnitude1}
  A=(1.3\pm 0.6)\times 10^{-3},~~~~B=(2.0\pm 2.2)\times 10^{-4}.
\end{equation}
We can see that the magnitude of the monopole is one order of magnitude smaller than that of the dipole. On the other hand, the statistical significance of the dipole is about at the $2\sigma$ level, while the statistical significance of the monopole is less than $1\sigma$. In fact, if we fit the data with a dipole only, we get a similar direction, pointing towards
\begin{equation}\label{eq:direccion2}
  (l,b)=(309.0^{\circ}\pm 22.4^{\circ},-19.3^{\circ}\pm 12.9^{\circ}),
\end{equation}
and the magnitude of the dipole is also about at the $2\sigma$ level, i.e.,
\begin{equation}\label{eq:magnitude2}
  A=(1.0\pm 0.5)\times 10^{-3}.
\end{equation}

It is interesting that the dipole direction derived from the DF method is close to the direction of the faintest SNe Ia derived from the HC method. The angle between these two directions is about $\Delta\theta\approx 33^{\circ}$. Therefore, some authors believe that the result of the DF method is coincident with that of the HC method \citep{Antoniou:2010,Mariano:2012}. However, an important thing has been ignored. Since the magnitudes of the dipole and monopole are positive, at the dipole direction, $\cos\theta=0$, and the right-hand-side of Eq.(\ref{eq:dipole}) is positive. Therefore, the right-hand-side of Eq.(\ref{eq:deviation-of-modulus}) is also positive, i.e., $\bar{\mu}>\mu$. This means that the observed distance modulus (so the luminosity distance) is smaller than that predicted by the ${\rm \Lambda}$CDM model. Thus, in the dipole direction, $\Omega_M$ is the largest (larger than the best-fit value of the ${\rm \Lambda}$CDM model), since $\mu$ decreases as $\Omega_M$ increases. This further implies that the dipole direction is the direction of the brightest SNe Ia. On the contrary, the faintest SNe Ia are in the opposite direction. This conflicts with the result of the HC method -- the direction of the faintest SNe Ia derived from the DF method is almost opposite to that from the HC method! In the following, we will show that the discrepancies between these two methods can be seen more clearly if the data points are uniformly distributed in the sky.

\subsection{Monte Carlo simulations}\label{sec:simulation}

This subsection aims to further verifying the discrepancies between the HC method and DF method. To address this issue, we perform Monte Carlo simulations. We generate a dataset which exactly follows the dipole distribution, and then perform the HC method on the dataset to see if the obtained preferred direction is close to the dipole direction or not. The steps of the simulations are given as follows:
\begin{enumerate}
  \item{Replace the positions of all SNe Ia with random positions, such that the SNe Ia are uniformly distributed in the sky. This can be done by replacing the coordinates of the $i$-th data point ($l_i,b_i$) with a pair of random numbers ($l,b$), where $l\in[0^{\circ}, 360^{\circ})$ and $b\in[-90^{\circ},90^{\circ}]$ are random numbers following the uniform distribution.}
  \item{The distance modulus of each supernova is replaced by the simulated distance modulus, calculated from $\mu_{\rm sim}(z)=\bar{\mu}(z)[1-A\cos\theta]$, where $\theta$ is the angle of the supernova with respect to the dipole direction given by Eq.(\ref{eq:direccion2}), and $A$ is the magnitude of the dipole given by Eq.(\ref{eq:magnitude2}). We endow each data the same weight, i.e., the uncertainties of all the data points are assumed to be the same. Therefore, the simulated data points always follow the dipole distribution, and performing the DF method on the simulated dataset always leads to the dipole direction $(l,b)=(309.0^{\circ}, -19.3^{\circ})$.}
  \item{Find the direction of the faintest SNe Ia (i.e., the direction of minimum $\Omega_M$) for the simulated dataset using the HC method described in section \ref{sec:hemisphere}.}
  \item{Repeat steps (i) to (iii) $n$ times (for example, $n=200$), we can find $n$ directions of the faintest SNe Ia, as well as $n$ magnitudes of anisotropy.}
\end{enumerate}

\begin{figure}
\centering
  \includegraphics[width=16 cm]{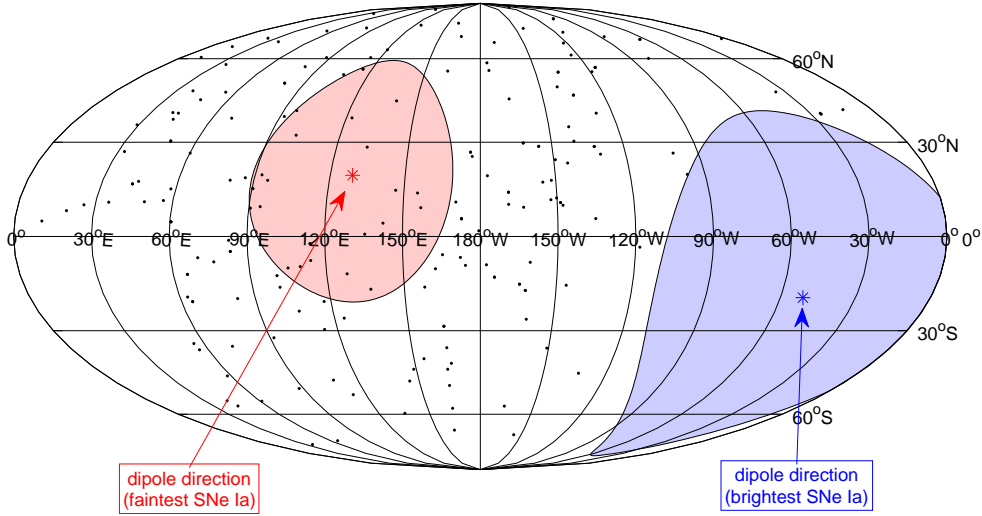}
  \vspace{-1.0cm}
  \caption{\small{The distribution of the faintest SNe Ia directions derived from the HC method in the 200 simulations. The blue star is the dipole direction (the brightest SNe Ia direction) pointing towards $(l,b)=(309.0^{\circ}, -19.3^{\circ})$, and the blue patch is a circular region with ``\,radius" $\Delta\theta\leq 60^{\circ}$ with respect to this direction. The red star is the opposite to the dipole direction (the faintest SNe Ia direction) pointing towards $(l,b)=(129.0^{\circ}, 19.3^{\circ})$, and the red patch is a circular region with ``\,radius" $\Delta\theta\leq 40^{\circ}$ with respect to this direction.}}\label{fig:direction}
\end{figure}

Fig.\ref{fig:direction} shows the obtained directions of the faintest SNe Ia in 200 simulations.
In each simulation, we choose 500 axes randomly in the hemisphere. The blue star is the dipole direction (i.e., the direction of the brightest SNe Ia derived from the DF method) pointing towards $(l,b)=(309.0^{\circ},-19.3^{\circ})$, and the red star is its opposite direction (i.e., direction of the faintest SNe Ia derived from the DF method). The blue patch is a circular region with ``\,radius" $\theta\leq 60^{\circ}$ with respective to the brightest SNe Ia direction, while the red patch is a circular region with ``\,radius" $\theta\leq 40^{\circ}$ with respective to the faintest SNe Ia direction. The black dots are the faintest SNe Ia directions derived from the HC method. If the HC method and the DF method are consistent, the black dots should cluster near the red star. However, we can see in Fig.\ref{fig:direction} that the black dots are almost uniformly distributed in the sky, except for the blue region. None of the black dots fall into the blue region. About 13.5\% (27 out of 200) black dots fall into the red region. This implies that the HC method and the DF method are inherently different methods and they can lead to very different results in many cases. There are only about 13.5\% possibility that they can get similar results (within $40^{\circ}$ uncertainty). It is unlikely that these two methods get two completely opposite directions (within $60^{\circ}$ uncertainty), if the dataset are uniformly distributed in the sky. The coincidence that the faintest SNe Ia directions derived from the HC method and that from the DF method are approximately opposite for the Union2 may be due to the nonuniform distribution of the dataset.

\begin{figure}
\centering
  \includegraphics[width=15 cm]{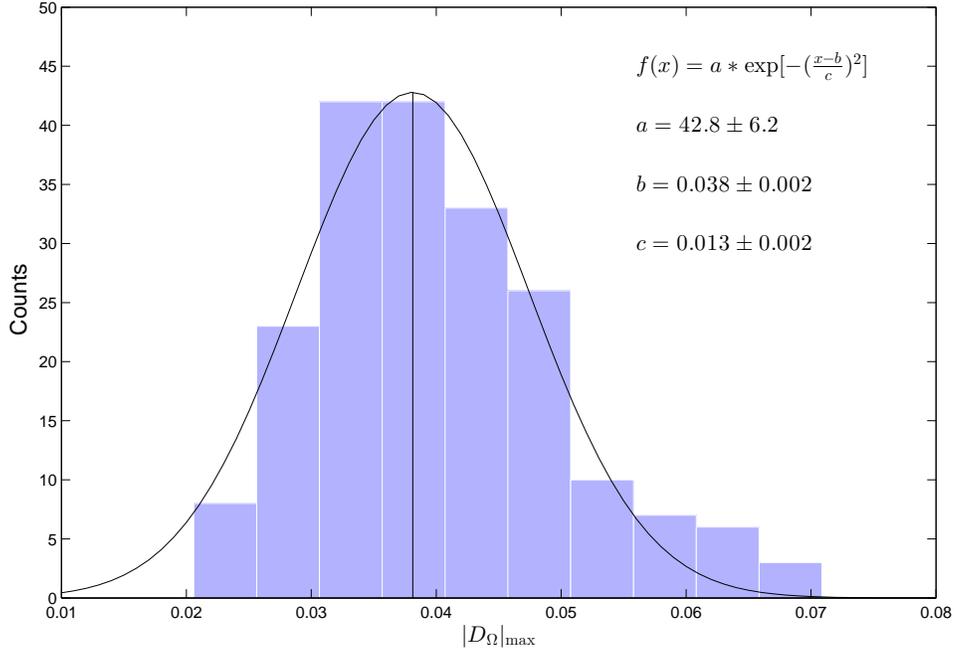}
  \vspace{-0.6cm}
  \caption{\small{The histogram of the maximum anisotropy of $\Omega_M$, (i.e., $|D_{\Omega}|_{\rm max}$) in the 200 simulations. It can be fitted well by the Gauss function, with the average value $\langle{|D_{\Omega}|_{\rm max}}\rangle=0.038$, and standard variance $\sigma=0.009$.}}\label{fig:omegam2}
\end{figure}

The maximum anisotropy of $\Omega_M$, (i.e., $|D_{\Omega}|_{\rm max}$) in the 200 simulations follows the Gauss distribution. The histogram of $|D_{\Omega}|_{\rm max}$ is plotted in Fig.\ref{fig:omegam2}.
It can be well fitted by the Gauss function
\begin{equation}
  f(|D_{\Omega}|_{\rm max})=a\exp\left[-\left(\frac{|D_{\Omega}|_{\rm max}-b}{c}\right)^2\right],
\end{equation}
with the best-fit values and their $1\sigma$ uncertainties
\begin{equation}
  a=42.8\pm 6.2,~~b=0.038\pm 0.002,~~c=0.013\pm 0.002.
\end{equation}
The average maximum anisotropy of the simulated data, $\langle{|D_{\Omega}|_{\rm max}}\rangle=0.038$, is about one order of magnitude smaller than that obtained from the Union2 dataset. This shows that both the magnitude of anisotropy and the preferred direction derived from the HC method strongly depend on the distribution of the data points, even though implementing the DF method to the simulated dataset always leads to the same magnitude and direction. The ability of the HC method to detect a signal in the dataset depends sensitively on the distribution of data points in the sky. The distribution-dependence of some methods in searching for the signal of anisotropy has already been noticed by \citet{Appleby:2014}.

\section{Discussion}\label{sec:discussion}

\begin{figure}
\centering
  \includegraphics[width=16 cm]{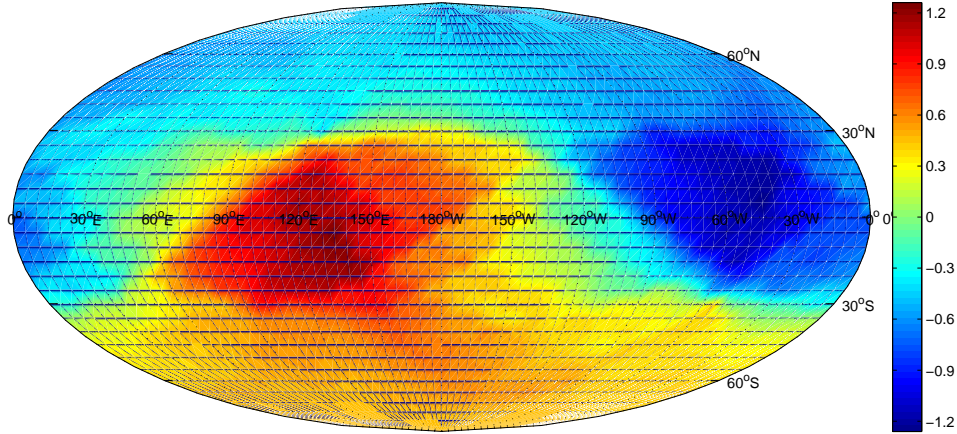}
  \vspace{-1.5cm}
  \caption{\small{The pseudo color map of $D_N(l,b)\equiv\Delta N/\bar{N}$. The most sparse direction points towards $(l,b)=(312^{\circ}, 2^{\circ})$, while the most clustered direction is its opposite.}}\label{fig:number}
\end{figure}

As we have showed, the faintest SNe Ia direction obtained from the HC method strongly depends on the distribution of the data points in the sky. Different distributions will lead to very different directions of the faintest SNe Ia, even if the data points are exactly follow the dipole distribution. To further show the relation between the faintest SNe Ia direction and the nonuniform distribution of the Union2 dataset, we perform a procedure similar to the HC method, but we take the number difference of SNe Ia in two opposite hemispheres as a diagnostic to quantify the anisotropy level of the distribution. Define
\begin{equation}
  D_N(l,b)\equiv\frac{\Delta N}{\bar{N}}=2\frac{N_u-N_d}{N_u+N_d},
\end{equation}
where $N_u$ and $N_d$ are the numbers of SNe Ia in the ``\,up" and ``\,down" hemispheres, respectively. We divide the sky into $5^{\circ}\times 5^{\circ}$ grids, and calculate the $D_N(l,b)$ value in the center of each grid. The pseudo color map of $D_N(l,b)$ is plotted in Fig.\ref{fig:number}.
As is showed, the most sparse direction points towards $(l,b)=(312^{\circ},2^{\circ})$, while the most clustered direction is its opposite direction, i.e., $(l,b)=(132^{\circ},-2^{\circ})$. The numbers of SNe Ia in these two hemispheres are 101 and 456, respectively. In these two directions, the maximum number anisotropy is $|D_N|_{\rm max}=1.27$. This implies that the distribution of the Union2 dataset in extremely nonuniform. Comparing the pseudo color maps of Fig.\ref{fig:omegam} and Fig.\ref{fig:number}, we can see that they are similar. Especially, it is amazing that the faintest SNe Ia direction derived from the HC method is so close to the most sparse direction (with angle difference $\Delta\theta\approx 16^{\circ}$), while the brightest SNe Ia direction is near the most clustered direction. The nonuniform distribution of the data points may significantly affect the detected signal of anisotropy. In addition, we just use the diagonal errors in the fitting. Considering the non-diagonal components of the covariance matrix may much increase the uncertainties, such that the signal of anisotropy even vanishes \citep{Jimenez:2014}.

There is another issue that should be payed specific attention when using the HC method. In order to search for the maximum anisotropy direction of the data points, we can choose an axis randomly and calculate the anisotropy level in this direction. If the axis runs over the sky, the maximum anisotropy direction can be found. Besides the random algorithm, we can divide the whole sky into equal longitude and latitude grids. Alternatively, we can divide the sky into equal area patches using the Healpix method \citep{Gorski:2005}. Since the random algorithm is the fast, it is used by most authors. In this paper, we also take the random algorithm to search for the preferred directions. But Fig.\ref{fig:omegam} and Fig.\ref{fig:number} are plotted using the equal longitude and latitude grids method, to make them seem more smooth. Some authors argue that it is enough to choose the number of random axes a little larger that the number of SNe Ia in each hemispheres, because change the direction of an axis does not change the corresponding $\Delta\Omega_M/\bar{\Omega}_M$ until a data point is crossed by the corresponding equator line \citep{Antoniou:2010,CaiTuo:2012}. ``\,Such a crossing is expected to occur when the direction of an axis changes by approximately the mean angular separation between data points" \citep{Antoniou:2010}. In fact, this is a misunderstanding. We will show in the following that much more axes should be chosen in order to ensure that only one data point is crossed when changing an axis to its adjacent one, although increasing the number of axes may not improve the accuracy significantly.

\begin{figure}
\centering
  \includegraphics[width=8 cm]{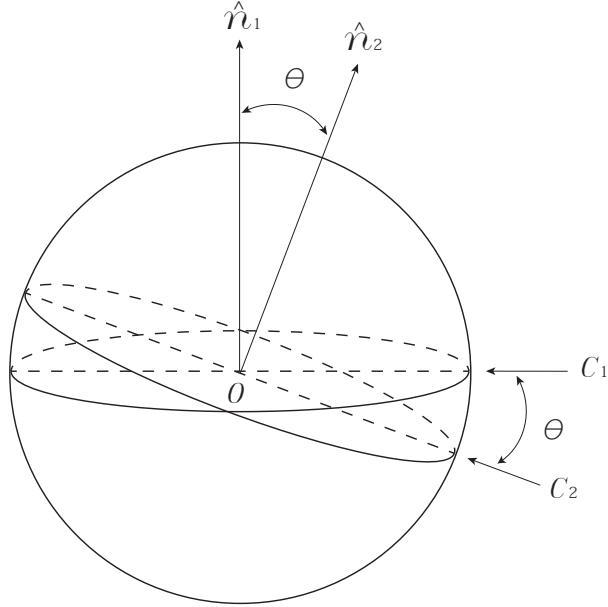}
  \vspace{0.4cm}
  \caption{\small{The celestial sphere. $\hat{\mathbf{n}}_1$ and $\hat{\mathbf{n}}_2$} are two directions with separation $\theta$, while $C_1$ and $C_2$ are the corresponding equators.}\label{fig:sphere}
\end{figure}

Suppose an axis is changed from $\hat{\mathbf{n}}_1$ to $\hat{\mathbf{n}}_2$, and the corresponding equator line is changed from $C_1$ to $C_2$ (see Fig.\ref{fig:sphere}).
The angle between $\hat{\mathbf{n}}_1$ and $\hat{\mathbf{n}}_2$ is denoted by $\theta$. The area of spherical surface swept by the equator is given as
\begin{equation}
  S=2\int_{\frac{\pi}{2}-\theta}^{\frac{\pi}{2}}\sin\theta d\theta\int_0^{\pi}d\varphi=2\pi\sin\theta.
\end{equation}
Assume that there are $N$ data points uniformly distributed in the sphere, then the number of data points swept by the equator is about
\begin{equation}
  \Delta N=\frac{S}{4\pi}N=\frac{N}{2}\sin\theta.
\end{equation}
If we randomly choose $n$ axes in one hemisphere, then the angle between two adjacent axes, $\Delta\theta$, can be approximately determined by
\begin{equation}
  \frac{2\pi}{n}=\int_0^{\Delta\theta}\sin\theta d\theta\int_0^{2\pi}d\varphi\approx\pi\Delta\theta^2,
\end{equation}
where we have assumed that $\Delta\theta\ll 1$. For $n=400$, as most authors take, we have $\Delta\theta\approx 4^{\circ}$. Thus, the number of data points crossed by the equator is about $\Delta N=N\sin\Delta\theta/2\approx 20$, where $N=557$ for the Union2 dataset. This is to say, if one axis is changed to its adjacent axis, there are about 10 data points go from the ``\,up" hemisphere to the ``\,down" hemisphere, and at the same time, about 10 data points go from the ``\,down" hemisphere to the ``\,up" hemisphere. This is the reason why the obtained $\Delta\Omega_M/\bar{\Omega}_M$ values are so sensitive to the directions. If we require that no more than one data point is crossed by the equator, the angle between two adjacent axes should be smaller than $\sim 0.4^{\circ}$. In other words, the axes in each hemisphere should be as much as $n\sim 4\times 10^4$, two order of magnitudes larger than the number of data points.

It should be mentioned that the Hubble parameter derived from the HC method is not exactly isotropic, although the magnitude of anisotropy is very small. It seems that $H_0$ is larger in the hemisphere where $\Omega_M$ is smaller. Even though a lower $\Omega_M$ implies fainter SNe Ia in this hemisphere, a higher value of $H_0$ indicates the opposite (i.e., brighter SNe Ia). If the contribution of $H_0$ is more significant than the contribution of $\Omega_M$, then this implies that the direction of lower $\Omega_M$ is also the direction of brighter SNe Ia. But this occurs only if the redshift is small enough. For large redshift, smaller $\Omega_M$ always leads to fainter SNe Ia.

\section{summary}\label{sec:summary}

In this paper, we compared the HC method and DF method in probing the anisotropic expansion of the Universe using the Union2 dataset. We found that the faintest SNe Ia directions obtained from the HC method and the DF method are approximately opposite. These two directions are either close to the most sparse direction of the data points, or near the most clustered direction. In order to further investigate the discrepancies between these two methods, we performed the Monte Carlo simulations. It was found that the faintest SNe Ia direction obtained from the HC method strongly depends on the distribution of the data points, although using the DF method we get a fixed direction. There is about 13.5\% probability that these two methods get the consistent directions, if the data points are uniformly distributed in the sky. It is unlikely that they lead to two completely opposite directions. The coincidence that the faintest SNe Ia directions obtained from these two method are approximately opposite in the Union2 dataset may be because of the extremely nonuniform distribution of the data points. We pointed out that choosing the number of axes a litter larger that the number of data points in each hemisphere is not accurate enough. The maximum accuracy can be achieved only if the number of axes is about two order of magnitudes larger than the number of data points in each hemisphere. However, even if we improve the accuracy, the conflictions between the HC method and DF method still can not be reconciled.

\section*{Acknowledgements}
We appreciate Prof. L. Perivolaropoulos for useful discussions. We are grateful to M.-H. Li, X. Li and S. Wang for meaningful suggestions. This work has been funded by the National Natural Science Fund of China under Grant No. 11375203.

\label{lastpage}

\end{document}